\documentclass[conference]{IEEEtran}
\IEEEoverridecommandlockouts %

\usepackage{cite} %
\usepackage{graphicx} %
\usepackage{amsmath,amsfonts,amsbsy,amssymb,amsthm}
\usepackage[caption=false,font=scriptsize]{subfig} %
\usepackage{url} %
\def\LongVersion{1}
\def\ShortVersion{2}
\def\MyVersion{\LongVersion}%

\def\StartLongVersion{\ifnum\MyVersion=\LongVersion}
\let\StopLongVersion=\fi
\def\StartShortVersion{\ifnum\MyVersion=\ShortVersion}
\let\StopShortVersion=\fi

\StartLongVersion
\def\headertext{\scriptsize~\hfill\thepage}
\def\footertext{}
\StopLongVersion

\StartShortVersion
\def\headertext{}
\def\footertext{}
\StopShortVersion

\makeatletter
\def\ps@headings{%
\def\@oddhead{\headertext}%
\def\@evenhead{\headertext}%
\def\@oddfoot{\footertext}%
\def\@evenfoot{\footertext}}
\makeatother
\usepackage[usenames]{color}
\DeclareMathAlphabet{\mathpzc}{OT1}{pzc}{m}{it}

\renewcommand{\binom}[2]{\left(\genfrac{}{}{0pt}{}{#1}{#2}\right)}

\newcommand{\BB}[2]{{\mathcal{B}\left(#1,#2\right)}}
\newcommand{\ceil}[1]{{\left\lceil{#1}\right\rceil}}
\newcommand{\ED}{{E_D}}
\newcommand{\EDm}[1]{{\mbox{$\ED\left(\lambda,T,m{=}#1\right)$}}}
\newcommand{\ee}{{\mathrm{e}}}
\newcommand{\EE}[1]{{\mathbb{E}\left[#1\right]}}

\newcommand{\floor}[1]{{\left\lfloor{#1}\right\rfloor}}
\newcommand{\II}[1]{{\mathbf{I}\left[#1\right]}}
\newcommand{\nset}{{\mbox{$\left\{1,\ldots,n\right\}$}}}
\newcommand{\PP}[1]{{\mathbb{P}\left[#1\right]}}

\newcommand{\Ps}{{P_{\text{S}}}}

\newcommand{\rr}{{\mathbf{r}}}

\newcommand{\xxs}{{\bar{\mathbf{x}}}}
\newcommand{\xxsm}[1]{{\mbox{$\xxs\left(n,T,m{=}#1\right)$}}}

\newcommand{\xxn}{{\mbox{$\left(x_1,\ldots,x_n\right)$}}}
\newcommand{\ZZ}{{\mathbb{Z}}}

\newcommand{\startcompact}[1]{\par\vspace{-0.75em}\begin{#1}\allowdisplaybreaks\ignorespaces}
\newcommand{\stopcompact}[1]{\end{#1}\ignorespaces}

\allowdisplaybreaks

\theoremstyle{definition}
\newtheorem*{thm:problem}{Problem}
\newtheorem*{thm:definition}{Definition}
\newtheorem*{thm:conjecture}{Conjecture}

\theoremstyle{plain}
\newtheorem{thm:claim}{Claim}
\newtheorem{thm:proposition}{Proposition}
\newtheorem{thm:lemma}{Lemma}
\newtheorem{thm:corollary}{Corollary}
\newtheorem{thm:theorem}{Theorem}

\begin{document}

\title{Distributed Storage Allocations for Optimal Delay
\thanks{\hrule width 0.25\columnwidth \vskip5pt
This work has been supported in part by the Air Force Office of Scientific Research under grant FA9550-10-1-0166 and Caltech's Lee Center for Advanced Networking.}
}%

\author{%
\IEEEauthorblockN{Derek Leong}
\IEEEauthorblockA{Department of Electrical Engineering\\
California Institute of Technology\\
Pasadena, California 91125, USA\\
\textit{derekleong@caltech.edu}}
\and
\IEEEauthorblockN{Alexandros G. Dimakis}
\IEEEauthorblockA{Department of Electrical Engineering\\
University of Southern California\\
Los Angeles, California 90089, USA\\
\textit{dimakis@usc.edu}}
\and
\IEEEauthorblockN{Tracey Ho}
\IEEEauthorblockA{Department of Electrical Engineering\\
California Institute of Technology\\
Pasadena, California 91125, USA\\
\textit{tho@caltech.edu}}}

\maketitle

\begin{abstract}
We examine the problem of creating an encoded distributed storage representation of a data object for a network of mobile storage nodes so as to achieve the optimal recovery delay.
A source node creates a single data object and disseminates an encoded representation of it to other nodes for storage, subject to a given total storage budget.
A data collector node subsequently attempts to recover the original data object by contacting other nodes and accessing the data stored in them.
By using an appropriate code, successful recovery is achieved when the total amount of data accessed is at least the size of the original data object.
The goal is to find an allocation of the given budget over the nodes that optimizes the recovery delay incurred by the data collector;
two objectives are considered:
(i) maximization of the probability of successful recovery by a given deadline, and
(ii) minimization of the expected recovery delay.
We solve the problem completely for the second objective in the case of \emph{symmetric} allocations (in which all nonempty nodes store the same amount of data), and show that the optimal symmetric allocation for the two objectives can be quite different.
A simple data dissemination and storage protocol for a mobile delay-tolerant network is evaluated under various scenarios via simulations.
Our results show that the choice of storage allocation can have a significant impact on the recovery delay performance, and that coding may or may not be beneficial depending on the circumstances.
\end{abstract}

\section{Introduction}

Consider a network of $n$ mobile storage nodes.
A source node creates a single data object of unit size (without loss of generality), and disseminates an encoded representation of it to other nodes for storage, subject to a given total storage budget $T$.
Let $x_i$ be the amount of coded data eventually stored in node \mbox{$i\in\nset$} at the end of the data dissemination process.
Any amount of data may be stored in each node, as long as the total amount of storage used over all nodes is at most the given budget $T$, that is, \mbox{$\sum_{i=1}^{n} x_i \leq T$}.

\begin{figure}
\centering
\includegraphics[width=0.5\columnwidth]{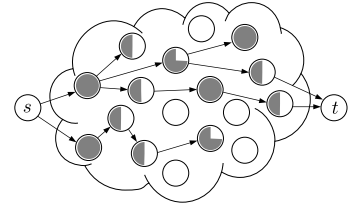}
\caption{Information flows originating at the source $s$, some of which finally arrive at the data collector $t$.
Different amounts of coded data may eventually be stored in each storage node, subject to the given total storage budget $T$.}
\label{fig:Network}
\end{figure}

At some time after the completion of the data dissemination process, a data collector node begins to recover the original data object by contacting other nodes and accessing the data stored in them.
We make the simplifying assumption that the stored data is instantaneously transmitted on contact;
this approximates the case where there is sufficient bandwidth and time for data transmission during each contact.
This data recovery process continues until the data object can be recovered from the cumulatively accessed data.
Let random variable $D$ denote the recovery delay incurred by the data collector, defined as the earliest time at which successful recovery can occur, measured from the beginning of the data recovery process.
\StartLongVersion
Fig.~\ref{fig:Network} depicts the information flows in such a network.
\StopLongVersion

\StartLongVersion
\StopLongVersion

By using an appropriate code for the data dissemination process and eventual storage, successful recovery can be achieved when the total amount of data accessed by the data collector is at least the size of the original data object.
This can be accomplished with random linear codes \cite{dl:ho06random,dl:fragouli06network} or a suitable MDS code, for example.
Thus, if \mbox{$\rr_d\subseteq\{1,\ldots,n\}$} is the set of all nodes contacted by the data collector by time $d$, then the recovery delay $D$ can be written as
\[
D \triangleq \min\left\{d:\sum_{i\in\rr_d} x_i\geq 1\right\}.
\]
Our goal is to find a storage allocation $\xxn$ that produces the optimal recovery delay, subject to the given budget constraint.
Specifically, we shall examine the following two objectives involving the recovery delay $D$:
\begin{enumerate}
\item
maximization of the probability of successful recovery by a given deadline $d$, or \textbf{recovery probability} \mbox{$\PP{D\leq d}$}, and
\item
minimization of the \textbf{expected recovery delay} \mbox{$\EE{D}$}.
\end{enumerate}
By solving for the optimal allocation, we will also be able to determine whether coding is beneficial for recovery delay.
For example, uncoded replication would suffice if each nonempty node is to store the data object in its entirety (i.e.~\mbox{$x_i\geq 1$} for all \mbox{$i\in S$}, and \mbox{$x_i=0$} for all \mbox{$i\notin S$}, where $S$ is some subset of \nset);
the data collector would not need to combine data accessed from different nodes in order to recover the data object.

The nodes of the network are assumed to move around and contact each other according to an exogenous random process;
they are unable to change their trajectories in response to the data dissemination or recovery processes.
(The recovery delay could be improved significantly if nodes were otherwise allowed to act on oracular knowledge about future contact opportunities \cite{dl:jain04routing}, for example.)

Most work on delay-tolerant networking traditionally assume that the data object is intended for immediate consumption;
both the data dissemination and recovery processes would therefore begin at the same time, and the recovery delay would be measured from the beginning of the data dissemination process.
In contrast, our model more accurately reflects the characteristics of longer-term storage where the data object can be consumed long after its creation.
Nonetheless, our model can still be a good approximation for short-term storage especially when the data dissemination process occurs very rapidly, as in the case of binary \textsc{spray-and-wait}  \cite{dl:spyropoulos05spray} where the number of nodes disseminating or spraying data grows exponentially over time.

We also note that in most of the literature involving distributed storage, either the data object is assumed to be replicated in its entirety (see, for e.g., \cite{dl:spyropoulos05spray}), or, if coding is used, every node is assumed to store the same amount of coded data (see, for e.g., \cite{dl:acedanski05how,dl:dimakis05ubiquitous,dl:kamra06growth,dl:lin07data,dl:aly08fountain}).
Allocations of a storage budget with nodes possibly storing different amounts of data are not usually considered.

\subsection{Our Contribution}

This paper attempts to address the gaps in our understanding of how the choice of storage allocation can affect the recovery delay performance.
We formulate a simple analytical model of the problem and show that the maximization of the \textbf{recovery probability} \mbox{$\PP{D\leq d}$} can be expressed in terms of the reliability maximization problem introduced in \cite{dl:karppersonal}.
It turns out that the simple strategies of spreading the budget minimally (i.e.~uncoded replication) and spreading the budget maximally over all $n$ nodes (i.e.~assigning \mbox{$x_i=\frac{T}{n}$} for all $i$) may both be suboptimal;
in fact, the optimal allocation may not even be symmetric
(we say that an allocation is \emph{symmetric} when all nonzero $x_i$ are equal).
Applying our earlier results \cite{dl:leong10symmetric}, we can show that minimal spreading is optimal among symmetric allocations when the deadline $d$ is sufficiently small, while maximal spreading is optimal among symmetric allocations when the deadline $d$ is sufficiently large.

For the minimization of the \textbf{expected recovery delay} \mbox{$\EE{D}$}, we are able to characterize the optimal \emph{symmetric} allocation completely:
minimal spreading (i.e.~uncoded replication) turns out to be optimal whenever the budget $T$ is an integer;
otherwise, the amount of spreading in the optimal symmetric allocation increases with the fractional part of $T$.

Interestingly, our analytical results demonstrate that the optimal symmetric allocation for the two objectives can be quite different.
In particular, when the budget $T$ is an integer, we observe a phase transition in the optimal symmetric allocation as the deadline $d$ increases, for the maximization of \textbf{recovery probability} \mbox{$\PP{D\leq d}$};
however, minimal spreading (i.e.~uncoded replication) alone turns out to be optimal for the minimization of \textbf{expected recovery delay} \mbox{$\EE{D}$}.

We proceed to apply our theoretical insights to the design of a simple data dissemination and storage protocol for a mobile delay-tolerant network.
Our protocol generalizes \textsc{spray-and-wait} \cite{dl:spyropoulos05spray} by allowing the use of variable-size coded packets.
Using network simulations, we compare the performance of different symmetric allocations under various circumstances.
These simulations allow us to capture the transient dynamics of the data dissemination process that were simplified in the analytical model.
Our main result shows that a maximal spreading of the budget is optimal in the \emph{high recovery probability regime}.
Specifically, maximal spreading can lead to a significant reduction in the wait time required to attain a desired recovery probability.
\StartLongVersion
We also evaluate the protocol against a real-world data set consisting of the mobility traces of taxi cabs operating in a city.
\StopLongVersion
Besides validating the predictions made in our theoretical analysis, these simulations also reveal several interesting properties of the allocations under different circumstances.

\subsection{Other Related Work}

Jain et al. \cite{dl:jain05using} and Wang et al. \cite{dl:wang05erasure} evaluated the delay performance of symmetric allocations experimentally in the context of routing in a delay-tolerant network.
Our results complement and generalize several aspects of their work.

We present a theoretical analysis of the problem in Section~\ref{sec:TheoreticalAnalysis}, and undertake a simulation study in Section~\ref{sec:SimulationStudy}.
\StartLongVersion
Proofs of theorems are deferred to the appendix.
\StopLongVersion
\StartShortVersion
Proofs of theorems can be found in the extended version of this paper \cite{dl:leong11delayarxiv}.
\StopShortVersion

\section{Theoretical Analysis}
\label{sec:TheoreticalAnalysis}

\noindent
We adopt the following notation throughout the paper:
\begin{tabbing}
\; \= $n$ \, \= total number of storage nodes, $n\geq 2$ \\
\> $\lambda$ \> contact rate between any given pair of  nodes, $\lambda>0$ \\
\> $x_i$ \> amount of data stored in node $i\in\nset$, $x_i \geq 0$ \\
\> $T$ \> total storage budget, $1 \leq T \leq n$ \\
\> $D$ \> random variable denoting recovery delay
\end{tabbing}
The indicator function is denoted by $\II{G}$, which equals $1$ if statement $G$ is true, and $0$ otherwise.
We use \mbox{$\BB{n}{p}$} to denote the binomial random variable with $n$ trials and success probability $p$.
An allocation $\xxn$ is said to be \emph{symmetric} when all nonzero $x_i$ are equal;
for brevity, let \mbox{$\xxs(n,T,m)$} denote the symmetric allocation for $n$ nodes that uses a total storage of $T$ and contains exactly \mbox{$m\in\nset$} nonempty nodes, that is,
\startcompact{small}
\begin{align*}
\xxs(n,T,m)
\triangleq
\biggl(\,
\underbrace{\frac{T}{m},\ldots,\frac{T}{m}}_{m \text{ terms}},
\underbrace{0,\ldots,0\vphantom{\frac{T}{m}}}_{(n-m) \text{ terms}}
\!\!\biggr).
\end{align*}
\stopcompact{small}

The number of contacts between any given pair of nodes in the network is assumed to follow a Poisson distribution with rate parameter $\lambda$;
the time between contacts is therefore described by an exponential distribution with mean $\frac{1}{\lambda}$.
Let $W_1,\ldots,W_n$ be i.i.d.~random variables denoting the times at which the data collector first contacts node $1,\ldots,n$, respectively, where \mbox{$W_i\sim\text{Exponential}(\lambda)$}.

\subsection{Maximization of Recovery Probability $\PP{D\leq d}$}
\label{sec:TheoreticalAnalysisRecoveryProb}

Let the given recovery deadline be \mbox{$d>0$}, and let the subset of nodes contacted by the data collector by time $d$ be \mbox{$\rr\subseteq\nset$}.
Successful recovery occurs by time $d$ if and only if the total amount of data stored in the subset $\rr$ of nodes is at least 1.
In other words, the recovery delay $D$ is at most $d$ if and only if \mbox{$\sum_{i\in\rr} x_i \geq 1$}.
Since the data collector contacts each node by time $d$ independently with constant probability $p_{\lambda,d} $, given by
\[
p_{\lambda,d} \triangleq \PP{W\leq d} = F_{W}(d) =
1 - \ee^{-\lambda d},
\]
it follows that the probability of contacting exactly a subset $\rr$ of nodes by time $d$ is \mbox{$p_{\lambda,d}^{|\rr|} (1-p_{\lambda,d})^{n-|\rr|}$}.
The recovery probability $\PP{D\leq d}$ can therefore be obtained by summing over all possible subsets $\rr$ that allow successful recovery:
\startcompact{small}
\begin{align}
\PP{D\leq d}
= \hspace{-1em} \sum_{\substack{\rr\subseteq\nset:\\|\rr|\geq 1}} \hspace{-1em}
p_{\lambda,d}^{|\rr|} (1-p_{\lambda,d})^{n-|\rr|}
\cdot \II{\sum_{i\in\rr} x_i \geq 1}.
\label{eq:RecoveryProb}
\end{align}
\stopcompact{small}
We seek an optimal allocation $\xxn$ of the budget $T$
(that is, subject to \mbox{$\sum_{i=1}^{n} x_i\leq T$}, where \mbox{$x_i\geq 0$} for all $i$)
that maximizes $\PP{D\leq d}$, for a given choice of $n$, $\lambda$, $d$, and $T$.

This problem matches the reliability maximization problem of \cite{dl:leong10symmetric} with $p_{\lambda,d}$ as the access probability;
we recall that the optimal allocation may be nonsymmetric and can be difficult to find.
However, if we restrict the optimization to only \emph{symmetric} allocations, then we can specify the solution for a wide range of parameter values of $p_{\lambda,d}$ and $T$.
Specifically, if $\lambda$ or $d$ is sufficiently small, e.g.~\mbox{$p_{\lambda,d}\leq\frac{1}{\ceil{T}}$}, then \xxsm{\floor{T}}, which corresponds to a minimal spreading of the budget (i.e.~uncoded replication), is an optimal symmetric allocation.
On the other hand, if $\lambda$ or $d$ is sufficiently large, e.g.~\mbox{$p_{\lambda,d}\geq\frac{4}{3\floor{T}}$}, then either \xxsm{\floor{\floor{\frac{n}{T}}T}} or \xxsm{n}, which correspond to a maximal spreading of the budget, is an optimal symmetric allocation.

\subsection{Minimization of Expected Recovery Delay $\EE{D}$}
\label{sec:TheoreticalAnalysisExpectedDelay}

Rewriting \eqref{eq:RecoveryProb} in terms of the underlying random variables gives us the following c.d.f. for the recovery delay $D$:
\startcompact{small}
\begin{align*}
F_D(t)
= \hspace{-1.5em} \sum_{\substack{\rr\subseteq\nset:\\|\rr|\geq 1}} \hspace{-1.5em}
\big(F_W(t)\big)^{|\rr|} \big(1-F_W(t)\big)^{n-|\rr|}
\!\cdot\! \II{\sum_{i\in\rr} x_i \geq 1}\!.
\end{align*}
\stopcompact{small}
Differentiating $F_D(t)$ wrt $t$ produces the p.d.f.
\startcompact{footnotesize}
\begin{align*}
f_D(t)
&= \hspace{-2em} \sum_{\substack{\rr\subseteq\nset:\\|\rr|\geq 1}} \hspace{-2em}
\big(F_W(t)\big)^{|\rr|-1}
\big(1-F_W(t)\big)^{n-|\rr|-1}
\big(|\rr|-n\,F_W(t)\big)
f_W(t)
\\*[-2.5em]
& \hspace{22em} \cdot \II{\sum_{i\in\rr} x_i \!\geq\! 1}\!.
\end{align*}
\stopcompact{footnotesize}
Therefore, assuming \mbox{$\sum_{i=1}^{n} x_i \geq 1$} which is necessary for successful recovery, we can compute the expected recovery delay as follows:
\startcompact{scriptsize}
\begin{align}
& \EE{D}
= \int_{0}^{\infty} t\,f_D(t)\;dt \notag
\\
&= \hspace{-1.5em} \sum_{\substack{\rr\subseteq\nset:\\|\rr|\geq 1}} \hspace{-1.1em}
\left( \hspace{-0.2em}
\int_{0}^{\infty} \hspace{-0.8em}
t \big(F_W\!(t)\big)^{|\rr|-1}
\big(1{-}F_W\!(t)\big)^{n-|\rr|-1}
\big(|\rr|{-}n\,F_W\!(t)\big)
f_W\!(t)
\,dt\hspace{-0.2em}\right) \notag
\\*[-2em]
& \hspace{28em} \cdot \II{\sum_{i\in\rr} x_i \geq 1} \notag
\\
&= \frac{1}{\lambda}
\left(
H_n -
\hspace{-0.5em} \sum_{\substack{\rr\subseteq\nset:\\1\leq|\rr|\leq n-1}}
\frac{1}{(n-|\rr|)\binom{n}{|\rr|}}
\cdot \II{\sum_{i\in\rr} x_i \geq 1} \right)\!, \label{eq:ExpectedDelay}
\end{align}
\stopcompact{scriptsize}
where \mbox{$H_n\triangleq\sum_{i=1}^{n} \frac{1}{i}$} is the $n^{\text{th}}$ harmonic number.
We seek an optimal allocation $\xxn$ of the budget $T$
(that is, subject to \mbox{$\sum_{i=1}^{n} x_i\leq T$}, where \mbox{$x_i\geq 0$} for all $i$)
that minimizes $\EE{D}$, for a given choice of $n$, $\lambda$, and $T$.
Note that the optimal allocation is independent of $\lambda$ for the minimization of $\EE{D}$ but not for the maximization of $\PP{D\leq d}$.

The optimal value of $\EE{D}$ can be bounded as follows:

\begin{thm:lemma}
\label{thm:lemma:ExpectedDelayLowerBound}
The expected recovery delay $\EE{D}$ of an optimal allocation is at least
\startcompact{small}
\begin{align*}
\frac{1}{\lambda} \left( H_n - \sum_{r=1}^{n-1} \frac{\min \left( \frac{rT}{n},1 \right)}{n-r} \right).
\end{align*}
\stopcompact{small}
\end{thm:lemma}

We make the following conjecture about the optimal allocation, based on our numerical observations:

\begin{thm:conjecture}
A \emph{symmetric} optimal allocation always exists for any $n$, $\lambda$, and $T$.
\end{thm:conjecture}

As a simplification, we now proceed to restrict the optimization to only \emph{symmetric} allocations (which are easier to describe and implement, and appear to perform well).
For the symmetric allocation $\xxs(n,T,m)$, successful recovery occurs by a given deadline $d$ if and only if
$\ceil{1\big\slash\left(\frac{T}{m}\right)}$
$=\ceil{\frac{m}{T}}$
or more nonempty nodes are contacted by the data collector by time $d$, out of a total of $m$ nonempty nodes.
It follows that the resulting recovery probability is given by
\mbox{$\PP{D\leq d}$} $=$ \mbox{$\PP{\BB{m}{p_{\lambda,d}}\geq\ceil{\frac{m}{T}}}$}.
We therefore obtain the following c.d.f. and p.d.f. for the recovery delay $D$:
\startcompact{small}
\begin{align*}
F_D(t)
&= \sum_{r=\ceil{\frac{m}{T}}}^{m} \binom{m}{r} \big(F_W(t)\big)^{r} \big(1-F_W(t)\big)^{m-r},
\\
f_D(t)
&= \binom{m}{\ceil{\frac{m}{T}}} \ceil{\frac{m}{T}}
\big(F_W(t)\big)^{\ceil{\frac{m}{T}}-1} \big(1{-}F_W(t)\big)^{m-\ceil{\frac{m}{T}}} f_W(t).
\end{align*}
\stopcompact{small}
Thus, we can compute the expected recovery delay as follows:
\begin{align*}
\EE{D}
\!\!=\!\! \int_{0}^{\infty} \!\!\!\! t\,f_D(t)\;dt
= \frac{1}{\lambda} \sum_{i=1}^{\ceil{\frac{m}{T}}} \frac{1}{m-\ceil{\frac{m}{T}}+i}
\triangleq
E_D(\lambda,T,m).
\end{align*}

\begin{figure}
\centering
\includegraphics[width=0.81\columnwidth]{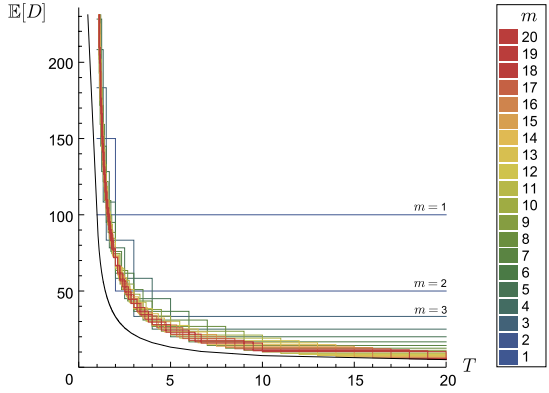}
\caption{%
Plot of expected recovery delay $\EE{D}$ against budget $T$ for each symmetric allocation $\xxs(n,T,m)$, for $(n,\lambda){=}\left(20,\frac{1}{100}\right)$.
Parameter $m$ denotes the number of nonempty nodes in the symmetric allocation.
The black curve gives a lower bound for the expected recovery delay of an optimal allocation, as derived in Lemma~\ref{thm:lemma:ExpectedDelayLowerBound}.}
\label{fig:ExpectedDelaySymmPlot}
\end{figure}

Fig.~\ref{fig:ExpectedDelaySymmPlot} compares the performance of different symmetric allocations over different budgets $T$, for an instance of $n$ and $\lambda$;
the value of $m$ corresponding to the optimal symmetric allocation appears to change in a nontrivial manner as we vary the budget $T$.
Fortunately, we can eliminate many candidates for the optimal value of $m$ by making the following observation (a similar observation was made in the maximization of the recovery probability \cite{dl:leong10symmetric}):
For fixed $n$, $\lambda$, and $T$, we have
\startcompact{small}
\begin{align*}
\ceil{\frac{m}{T}}
&= k \hspace{3.55em}
\text{when }
m\in \big((k-1)T,kT\big],
\intertext{for $k = 1,2,\ldots,\floor{\frac{n}{T}}$, and finally,}
\ceil{\frac{m}{T}}
&= \floor{\frac{n}{T}}+1
\text{ when }
m\in \left(\floor{\frac{n}{T}}T,n\right].
\end{align*}
\stopcompact{small}
Since
$\frac{1}{\lambda} \sum_{i=1}^{k} \frac{1}{m-k+i}$
is decreasing in $m$ for constant $\lambda$ and $k$, it follows that $\ED(\lambda,T,m)$ is minimized over each of these intervals of $m$ when we pick $m$ to be the largest integer in the corresponding interval.
Thus, given $n$, $\lambda$, and $T$, we can find an optimal $m^*$ that minimizes $\ED(\lambda,T,m)$ over all $m$ from among $\ceil{\frac{n}{T}}$ candidates:
\startcompact{small}
\begin{align}
\left\{\floor{T},\floor{2T},\ldots,\floor{\floor{\frac{n}{T}}T},n\right\}. \label{eq:CandidateMs}
\end{align}
\stopcompact{small}
Note that when \mbox{$m=\floor{kT}$}, \mbox{$k\in\ZZ^+$}, the expected recovery delay simplifies to the following expression:
\startcompact{small}
\begin{align*}
\EDm{\floor{kT}}
= \frac{1}{\lambda} \sum_{i=1}^{k} \frac{1}{\floor{kT}-k+i}.
\end{align*}
\stopcompact{small}

By further eliminating suboptimal candidate values for $m^*$ using suitable bounds for the harmonic number, we are able to completely characterize the optimal symmetric allocation for any $n$, $\lambda$, and $T$:

\begin{thm:theorem}
\label{thm:theorem:ExpectedDelayOptimalSymmetric}
Suppose \mbox{$T=a+1-\frac{1}{\ell}$}, where \mbox{$a\in\ZZ^+$}, \mbox{$\ell\geq 1$}.

\noindent
If \mbox{$\floor{\ell}\leq\floor{\frac{n}{T}}$}, then
\[
\xxsm{\floor{\floor{\ell}T}}
\]
is an optimal symmetric allocation;
if \mbox{$\floor{\ell}>\floor{\frac{n}{T}}$}, then
\[
\text{either } \xxsm{\floor{\floor{\frac{n}{T}}T}} \text{ or } \xxsm{n}
\]
is an optimal symmetric allocation.
\end{thm:theorem}

\noindent
If the budget $T$ is an integer (i.e.~\mbox{$\ell=1$}), then \mbox{$\floor{\ell}\leq\floor{\frac{n}{T}}$} is always true, and so \xxsm{\floor{T}}, which corresponds to a minimal spreading of the budget (i.e.~uncoded replication), is an optimal symmetric allocation.
However, if the budget $T$ is not an integer (i.e.~\mbox{$\ell>1$}), then the amount of spreading in the optimal symmetric allocation increases with the fractional part of $T$, up to a point at which either \xxsm{\floor{\floor{\frac{n}{T}}T}} or \xxsm{n}, which correspond to a maximal spreading of the budget, becomes optimal.
\StartLongVersion
Minimal spreading (i.e.~uncoded replication) therefore performs well over the whole range of budgets $T$, being optimal among symmetric allocations whenever $T$ is an integer
(its suboptimality at noninteger \mbox{$T=T_0$} can be bounded by the step difference in \EDm{\floor{T}} between \mbox{$T=T_0$} and \mbox{$T=\ceil{T_0}$}, since \mbox{$\ED(\lambda,T,m)$} is a nonincreasing function of $T$).
\StopLongVersion
\StartShortVersion
Minimal spreading (i.e.~uncoded replication) therefore performs well over the whole range of budgets $T$, being optimal among symmetric allocations whenever $T$ is an integer.
\StopShortVersion

In summary, we note that the optimal symmetric allocation for the two objectives can be quite different.
In particular, when the budget $T$ is an integer, we observe a phase transition from a regime where minimal spreading is optimal to a regime where maximal spreading is optimal, as the deadline $d$ increases, for the maximization of \textbf{recovery probability} \mbox{$\PP{D\leq d}$};
however, with the averaging over both regimes, minimal spreading (i.e.~uncoded replication) alone turns out to be optimal for the minimization of \textbf{expected recovery delay} \mbox{$\EE{D}$}.

\section{Simulation Study}
\label{sec:SimulationStudy}

\StartLongVersion
\begin{figure*}
\centering
\subfloat[Budget \mbox{$T=5$}]{\includegraphics[height=2.65in]{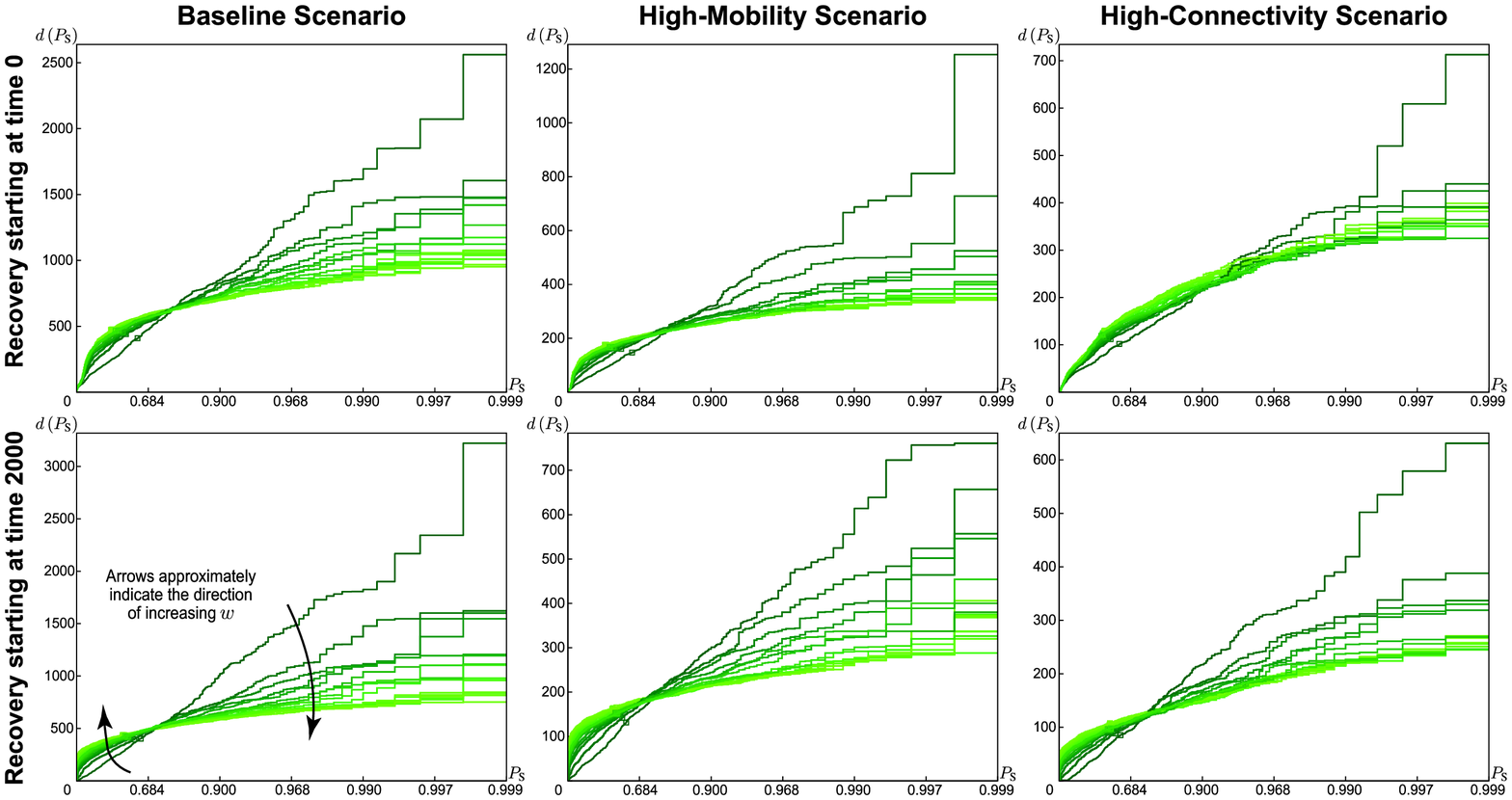}}
\\
\subfloat[Budget \mbox{$T=10$}]{\includegraphics[height=2.65in]{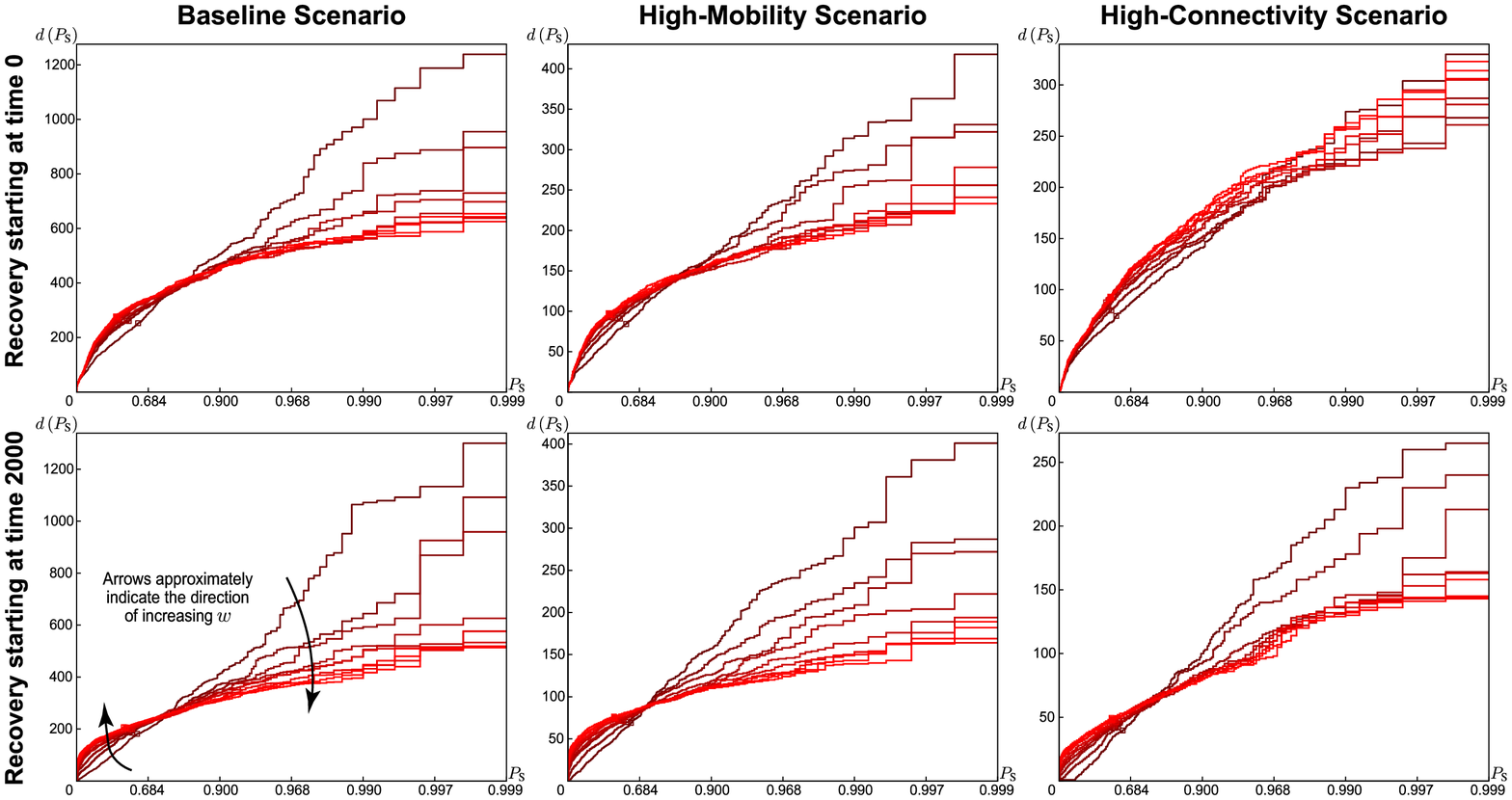}}
\\
\subfloat[Budget \mbox{$T=20$}]{\includegraphics[height=2.65in]{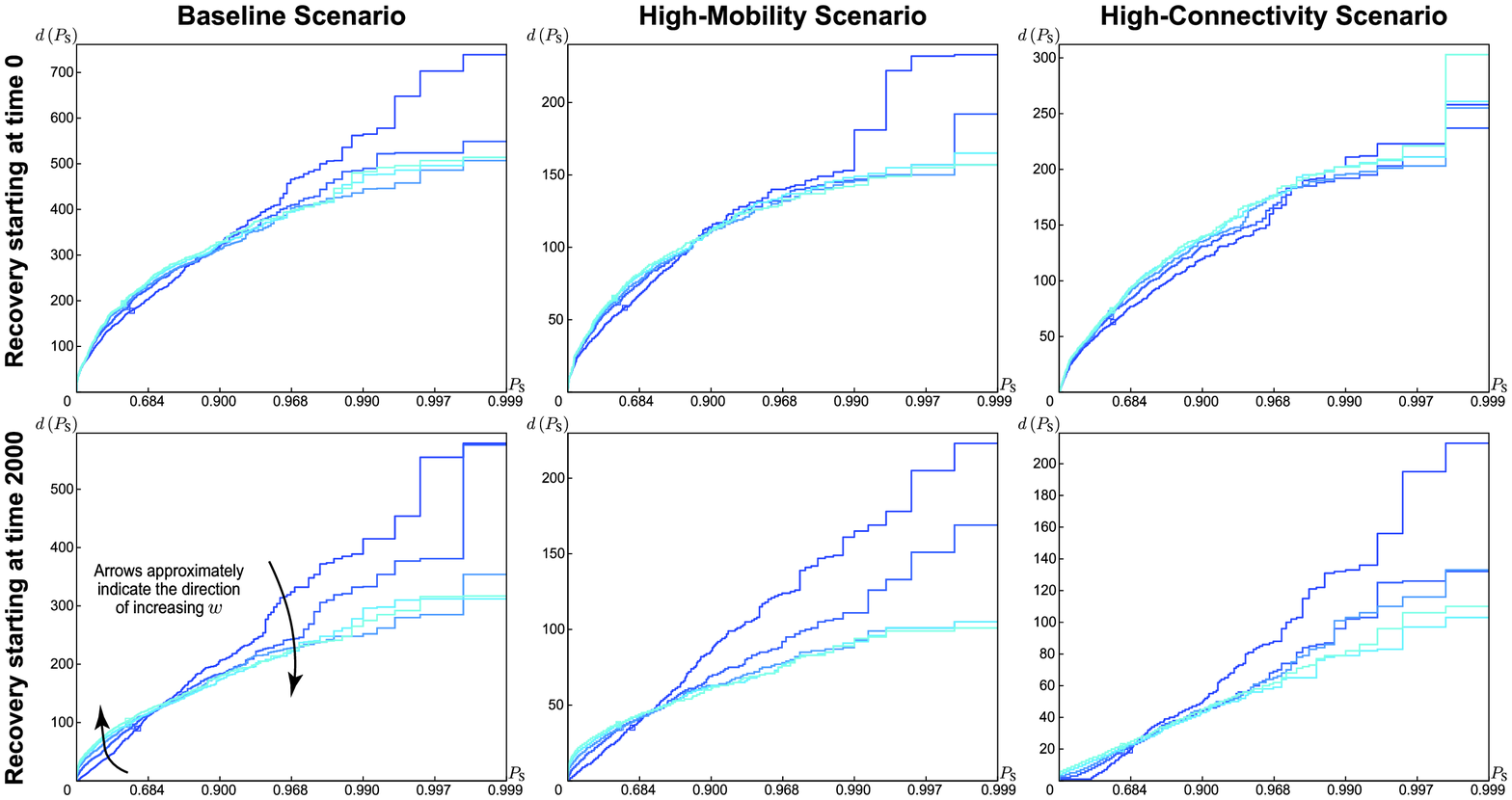}}
\caption{%
(Random Waypoint)
Plots of required wait time \mbox{$d(\Ps)$} against desired recovery probability $\Ps$ (semilogarithmic-scale), for budgets \mbox{$T=5,10,20$}.
Each colored line represents a specific choice of parameter \mbox{$w\in\left\{1,\ldots,\frac{n}{T}\right\}$}, with \mbox{$w=1$} (darkest) corresponding to a minimal spreading of the budget (i.e.~uncoded replication), and \mbox{$w=\frac{n}{T}$} (lightest) corresponding to a maximal spreading of the budget.
The mean recovery delay corresponding to each line is indicated by a square marker.}
\label{fig:SimPlots}
\end{figure*}
\begin{figure*}
\centering
\subfloat[Budget \mbox{$T=5$}]{\includegraphics[height=2.65in]{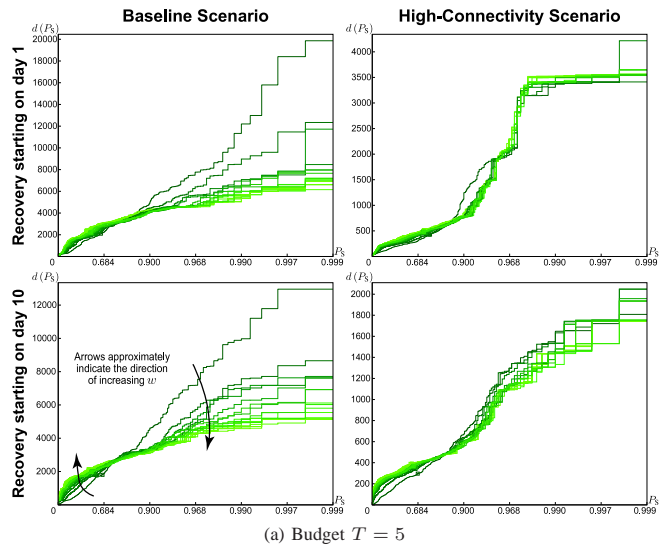}}
\\
\subfloat[Budget \mbox{$T=10$}]{\includegraphics[height=2.65in]{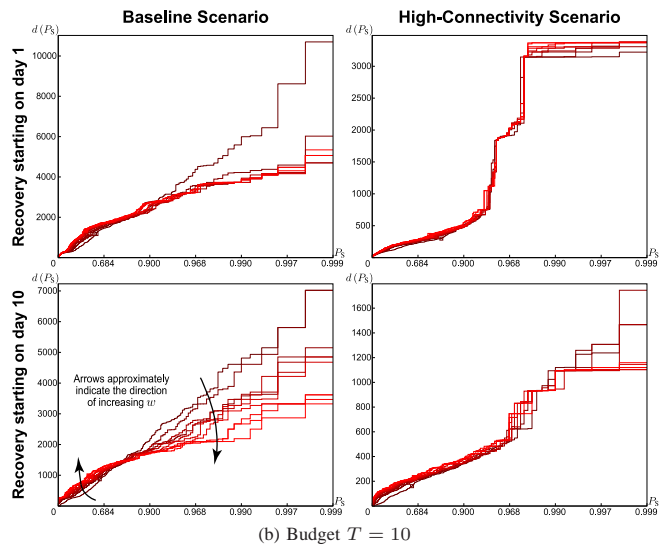}}
\\
\subfloat[Budget \mbox{$T=20$}]{\includegraphics[height=2.65in]{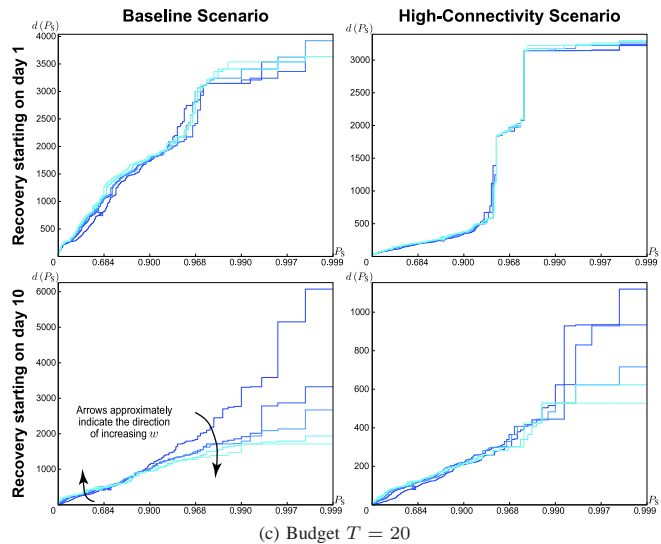}}
\caption{%
(Mobility Traces)
Plots of required wait time in minutes \mbox{$d(\Ps)$} against desired recovery probability $\Ps$ (semilogarithmic-scale), for budgets \mbox{$T=5,10,20$}.
Each colored line represents a specific choice of parameter \mbox{$w\in\left\{1,\ldots,\frac{n}{T}\right\}$}, with \mbox{$w=1$} (darkest) corresponding to a minimal spreading of the budget (i.e.~uncoded replication), and \mbox{$w=\frac{n}{T}$} (lightest) corresponding to a maximal spreading of the budget.
The mean recovery delay corresponding to each line is indicated by a square marker.}
\label{fig:CabsPlots}
\end{figure*}
\StopLongVersion

\StartShortVersion
\begin{figure*}
\centering
\includegraphics[height=3.03in]{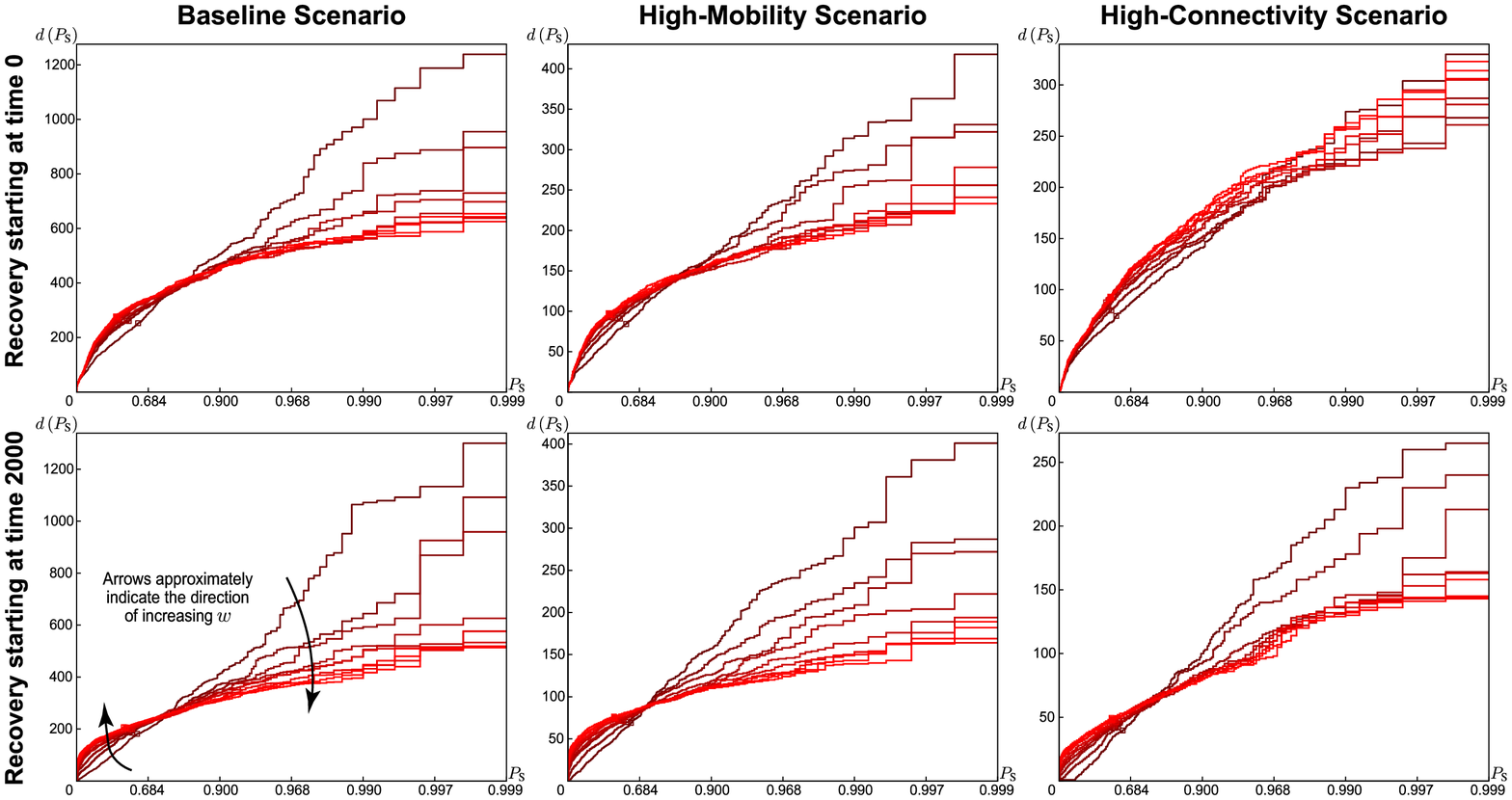}
\caption{%
Plots of required wait time \mbox{$d(\Ps)$} against desired recovery probability $\Ps$ (semilogarithmic-scale), for budget \mbox{$T=10$}.
Each colored line represents a specific choice of parameter \mbox{$w\in\left\{1,\ldots,\frac{n}{T}\right\}$}, with \mbox{$w=1$} (darkest) corresponding to a minimal spreading of the budget (i.e.~uncoded replication), and \mbox{$w=\frac{n}{T}=10$} (lightest) corresponding to a maximal spreading of the budget.
The mean recovery delay corresponding to each line is indicated by a square marker.}
\label{fig:SimPlotsT10}
\end{figure*}
\StopShortVersion

We apply our theoretical insights to the design of a simple data dissemination and storage protocol for a mobile delay-tolerant network.
Our protocol extends \textsc{spray-and-wait} \cite{dl:spyropoulos05spray} by allowing nodes to store \emph{coded} packets that are each $\frac{1}{w}$ the size of the original data object, where parameter $w$ is a positive integer;
successful recovery occurs when the data collector accesses at least $w$ such packets.
Different \emph{symmetric} allocations of the given total storage budget $T$ can be realized by choosing different values of $w$;
the original protocol, which uses uncoded replication, corresponds to \mbox{$w=1$}.

\subsection{Protocol Description}

The source node begins with a total storage budget of $T$ times the size of the original data object, which translates to $wT$ coded packets, each $\frac{1}{w}$ the size of the original data object.
Whenever a node with more than one packet contacts another node without any packets, the former gives \emph{half its packets} to the latter.
The actual amount of data stored or transmitted by a node never exceeds the size of the original data object (or $w$ packets) since the excess packets can always be generated on demand (using random linear coding, for example).
To reduce the total transmission cost incurred, a node can also directly transmit \emph{one packet} to each node it meets when it has $w$ or fewer packets left;
otherwise, these last few packets would be transmitted multiple times by different nodes.
The dissemination process is completed when no node has more than one packet.

\subsection{Network Model and Simulation Setup}

We implemented a discrete-time simulation of \mbox{$n=\text{100}$} wireless mobile nodes in a \mbox{1000$\times$1000} grid.
A random waypoint mobility model is assumed where at each time step, each node moves a random distance \mbox{$L\sim\text{Uniform}$[5,10]} towards a selected destination;
on arrival, the node selects a random point on the grid as its next destination.
Each node has a communication range of 20, and the bandwidth of each point-to-point link is large enough to support the transmission of $w$ packets in one time step.
At each time step, a maximal number of transmissions are randomly scheduled such that each node can transmit to or receive from at most one other node in range, and exactly one node may transmit in the range of a node receiving a transmission.
In addition to this \textbf{baseline scenario}, we also considered the following two scenarios:
\begin{enumerate}
\item
a \textbf{high-mobility scenario}, where the distance traveled by each node is increased to \mbox{$L\sim\text{Uniform}$[25,50]}, and
\item
a \textbf{high-connectivity scenario}, where the communication range is increased to 80.
\end{enumerate}
We measured the recovery delay incurred by the data collector for two cases:
\begin{enumerate}
\item
when the data recovery process begins at \textbf{time 0}, i.e.~at the beginning of the data dissemination process, and
\item
when the data recovery process begins at \textbf{time 2000}, i.e.~when the data dissemination process is already underway or completed.
(This is a more appropriate performance metric for longer-term storage.)
\end{enumerate}
We ran the simulation 500 times for each choice of budget \mbox{$T\in\{\text{5,10,20}\}$} and parameter \mbox{$w\in\left\{1,2,\ldots,\frac{n}{T}\right\}$} under each scenario, with a random pair of nodes appointed as the source and data collector for each run.

\subsection{Simulation Results}

\StartShortVersion

We briefly summarize our findings here;
detailed simulation results can be found in the extended version of this paper \cite{dl:leong11delayarxiv}.

Fig.~\ref{fig:SimPlotsT10} shows how the required wait time \mbox{$d(\Ps)$}, given by
\[
d(\Ps) \triangleq
\min\{d:\PP{D\leq d}\geq\Ps\},
\]
varies with the desired recovery probability $\Ps$ for each choice of parameter $w$, for budget \mbox{$T=10$};
these plots essentially describe how much time must elapse before a desired percentage of data collectors are able to recover the data object.
The phase transition predicted in the analytical model (Section~\ref{sec:TheoreticalAnalysisRecoveryProb}) is clearly discernible in all the plots, except for the high-connectivity scenario with recovery starting at time 0.
The mean recovery delay performance is also consistent with our analysis (Section~\ref{sec:TheoreticalAnalysisExpectedDelay}), with minimal spreading of the budget (\mbox{$w=1$}) being optimal.

We observe that in the \emph{high recovery probability regime}, maximal spreading of the budget (\mbox{$w=10$}) can lead to a significant reduction in the required wait time (by as much as 40\% to 60\% in the baseline and high-mobility scenarios).
We also note that the recovery start time appears to have a limited impact on the delay performance for the baseline and high-mobility scenarios:
for recovery starting at time 0, the different allocations yield about the same performance in the low recovery probability regime;
this can be explained by the similarity of the different allocations in the early stages of the dissemination process, when only a few nodes have been reached by the source directly or indirectly through relays.

\StopShortVersion

\StartLongVersion

Fig.~\ref{fig:SimPlots} shows how the required wait time \mbox{$d(\Ps)$}, given by
\[
d(\Ps)
\triangleq
\min\{d:\PP{D\leq d}\geq\Ps\},
\]
varies with the desired recovery probability $\Ps$ for each choice of parameter $w$;
these plots essentially describe how much time must elapse before a desired percentage of data collectors are able to recover the data object.
The \textbf{recovery probability} performance of the protocol (which can be inferred by flipping the axes) is mostly consistent with our analysis in Section~\ref{sec:TheoreticalAnalysisRecoveryProb};
specifically, the phase transition in the optimal symmetric allocation is clearly discernible in most of the plots.
The \textbf{expected recovery delay} performance is also mostly consistent with our analysis in Section~\ref{sec:TheoreticalAnalysisExpectedDelay}, with minimal spreading of the budget (\mbox{$w=1$}) being optimal in most of the plots.

The plots for the high-mobility scenario appear to be vertically scaled versions of the plots for the baseline scenario.
This is not surprising because an increase in node mobility approximately translates to a speeding up of time.
The effect of increasing node connectivity, on the other hand, seems less straightforward:
the phase transition in the optimal symmetric allocation is evident for recovery starting at time 2000 but not for recovery starting at time 0.
This discrepancy suggests that the data dissemination process is somewhat impeded by the increased connectivity, possibly due to greater interference.

We observe that in the \emph{high recovery probability regime}, maximal spreading of the budget (\mbox{$w=\frac{n}{T}$}) can lead to a significant reduction in the required wait time.
For example, given a budget of \mbox{$T=10$} and a desired recovery probability of \mbox{$\Ps=0.99$}, choosing maximal spreading (\mbox{$w=10$}) instead of minimal spreading or uncoded replication (\mbox{$w=1$}) can yield a reduction of 40\% to 60\% in the required wait time for the baseline and high-mobility scenarios.

We also observe that the recovery start time appears to have a limited impact on how the different allocations perform relative to each other;
the most noticeable effect of starting recovery at time 0 is the reduced spread in performance across different choices of parameter $w$, especially in the low recovery probability regime.
This can be explained by the similarity of the different allocations during the data dissemination process:
in the beginning, the different choices of parameter $w$ would see the same allocation of the budget over the nodes because only a few nodes have been reached by the source directly or indirectly through relays;
the different allocations are eventually realized only after a sufficient amount of time has passed.

\subsection{Evaluation on Mobility Traces}

To gain a better understanding of how our protocol might perform in a real-world setting, we evaluated it on a \mbox{CRAWDAD} data set comprising mobility traces of taxi cabs in San Francisco \cite{dl:piorkowski09parsimonious}.
The traces of 100 randomly selected cabs with GPS coordinate readings over the span of an 18-day period were used.
The GPS readings were sampled at approximately 60-second intervals;
because reading times were not synchronized across cabs, we estimated the position of a cab at any given time using linear interpolation.
For better accuracy, we assumed that a cab became inactive whenever the time between consecutive readings exceeded 2 minutes.
As in the preceding simulations, we considered different scenarios and data recovery start times.
Two scenarios were considered here:
\begin{enumerate}
\item
a \textbf{baseline scenario}, where the communication range of each cab is 20\,m, and
\item
a \textbf{high-connectivity scenario}, where the communication range is increased to 80\,m.
\end{enumerate}
We measured the recovery delay incurred by the data collector for two cases:
\begin{enumerate}
\item
when the data recovery process begins on \textbf{day 1}, and
\item
when the data recovery process begins on \textbf{day 10}, i.e.~half-way through the 18-day period.
\end{enumerate}
We ran the simulation 500 times for each choice of budget \mbox{$T\in\{\text{5,10,20}\}$} and parameter \mbox{$w\in\left\{1,2,\ldots,\frac{n}{T}\right\}$} under each scenario, with a random pair of cabs appointed as the source and data collector for each run.

Fig.~\ref{fig:CabsPlots} shows how the required wait time \mbox{$d(\Ps)$} varies with the desired recovery probability $\Ps$ for each choice of parameter $w$.
Compared to the plots of Fig.~\ref{fig:SimPlots} for the random waypoint simulations, these plots exhibit distinct ``jumps'' in the wait times, which can be attributed to the reduced mobility of the cabs at night.
Despite these nonideal conditions, many of the observations made for the previous simulations are still applicable here.
For instance, the phase transition in the optimal symmetric allocation is discernible in most of the plots for the baseline scenario.
Also, starting recovery on day 1 has the effect of reducing the spread in performance across different choices of parameter $w$, especially in the low recovery probability regime.

Once again, we observe that in the \emph{high recovery probability regime}, maximal spreading of the budget (\mbox{$w=\frac{n}{T}$}) can lead to a significant reduction in the required wait time.
For example, given a budget of \mbox{$T=10$} and a desired recovery probability of \mbox{$\Ps=0.99$}, choosing maximal spreading (\mbox{$w=10$}) instead of minimal spreading or uncoded replication (\mbox{$w=1$}) can yield a reduction of 30\% to 50\% in the required wait time for the baseline scenario.

\StopLongVersion

\StartLongVersion

\section{Conclusion}

We examined the recovery delay performance of different distributed storage allocations for a network of mobile storage nodes.
Our theoretical analysis and simulation study show that the choice of objective function (i.e.~recovery probability \textit{vs} expected recovery delay) can lead to very different optimal symmetric allocations, and that picking the right allocation for the given circumstances can make a significant difference in performance.

The work in this paper can be extended in several directions.
The simple contact model assumed here can be generalized to the case where a variable amount of data is transmitted during each contact between nodes.
Another natural generalization is to allow nonuniform contact rates $\lambda_i$ between the data collector and individual nodes.

\StopLongVersion

\StartLongVersion

\appendix[Proofs of Theorems]

\begin{proof}[Proof of Lemma~\ref{thm:lemma:ExpectedDelayLowerBound}]
Consider a feasible allocation \mbox{$\left(x_1,\ldots,x_n\right)$};
we have \mbox{$\sum_{i=1}^{n} x_i \leq T$}, where \mbox{$x_i\geq 0$}, \mbox{$i=1,\ldots,n$}.
Let $S_r$ denote the number of $r$-subsets of \mbox{$\left\{x_1,\ldots,x_n\right\}$} that have a sum of at least 1, where \mbox{$r\in\{1,\ldots,n\}$}.
Recall from Lemma~1 in \cite{dl:leong10symmetric} that $S_r$ can be bounded as follows:
\[
S_r \leq \min \left( \binom{n-1}{r-1}T, \binom{n}{r} \right).
\]
We can now rewrite \eqref{eq:ExpectedDelay} in  terms of $S_r$ by enumerating subsets according to size:
\begin{align*}
\EE{D}
&= \frac{1}{\lambda} \left( H_n - \sum_{r=1}^{n-1} S_r \cdot \frac{1}{(n-r)\binom{n}{r}} \right)
\\
&\geq \frac{1}{\lambda} \left( H_n - \sum_{r=1}^{n-1} \frac{\min \left( \binom{n-1}{r-1}T, \binom{n}{r} \right)}{(n-r)\binom{n}{r}} \right)
\\
&= \frac{1}{\lambda} \left( H_n - \sum_{r=1}^{n-1} \frac{\min \left( \frac{rT}{n},1 \right)}{n-r} \right).
\end{align*}
\hfill~
\end{proof}

\begin{proof}[Proof of Theorem~\ref{thm:theorem:ExpectedDelayOptimalSymmetric}]
Suppose $T=a+1-\frac{1}{\ell}$, where \mbox{$a\in\ZZ^+$}, \mbox{$\ell\geq 1$}.
Since \mbox{$kT=(a+1)k-\frac{k}{\ell}$}, the expected recovery delay for the symmetric allocation \xxsm{\floor{kT}}, where \mbox{$k\in\ZZ^+$}, can be written as
\begin{align*}
\EDm{\floor{kT}}
&= \frac{1}{\lambda} \sum_{i=1}^{k} \frac{1}{(a+1)k-\ceil{\frac{k}{\ell}}-k+i}
\\
&= \frac{1}{\lambda} \sum_{i=1}^{k} \frac{1}{ak-\ceil{\frac{k}{\ell}}+i}.
\end{align*}
Observe that \mbox{$\ceil{\frac{k}{\ell}}=v$} when \mbox{$k\in\big((v-1)\ell,v\ell\big]$}, for \mbox{$v=1,2,\ldots$}.
To compare \EDm{\floor{kT}} within each of these intervals of $k$, we introduce Lemma~\ref{thm:lemma:ExpectedDelaySymmetricFixedV}:

\begin{thm:lemma}
\label{thm:lemma:ExpectedDelaySymmetricFixedV}
For $a,v,k\in\ZZ^+$, $k\geq\frac{v}{a}$, the function
\[
f(a,v,k)
\triangleq \sum_{i=1}^{k} \frac{1}{ak-v+i}
= H_{ak-v+k} - H_{ak-v}
\]
decreases with $k$.
\end{thm:lemma}

\begin{proof}[Proof of Lemma~\ref{thm:lemma:ExpectedDelaySymmetricFixedV}]
Let $\Delta(a,v,k)$ denote the difference in the function value between consecutive values of $k$, that is,
\startcompact{small}
\begin{align*}
& \Delta(a,v,k) \triangleq f(a,v,k) - f(a,v,k+1)
\\
&= (H_{ak-v+k} - H_{ak-v}) - (H_{ak-v+k+a+1} - H_{ak-v+a})
\\
&= (H_{ak-v+a} - H_{ak-v}) - (H_{ak-v+k+a+1} - H_{ak-v+k})
\\
&=\left( \sum_{i=1}^{a} \frac{1}{ak-v+i} - \frac{1}{ak-v+k+i} \right) - \frac{1}{ak-v+k+a+1}
\\
&=\left( \sum_{i=1}^{a} \frac{k}{(ak-v+i)(ak-v+k+i)} \right) - \frac{1}{ak-v+k+a+1}.
\end{align*}
\stopcompact{small}
We will proceed to show that $\Delta(a,v,k)>0$ for any \mbox{$a,v,k\in\ZZ^+$}, \mbox{$k\geq\frac{v}{a}$}.
First, we find a lower bound for the summation term using a geometrical argument.
Consider the function
\[
g(t) \triangleq \frac{k}{(ak-v+t)(ak-v+k+t)},
\]
which has the second derivative
\[
g''(t) = \frac{2}{(ak-v+t)^{3}} - \frac{2}{(ak-v+k+t)^{3}}.
\]
For any \mbox{$a,v,k\in\ZZ^+$}, \mbox{$k\geq\frac{v}{a}$}, the function $g(t)$ is positive, decreasing with $t$, and convex (since \mbox{$g''(t)>0$}), on the interval \mbox{$t\in(0,\infty)$}.
We therefore have the lower bound
\startcompact{small}
\begin{align*}
\sum_{i=1}^{a} \frac{k}{(ak-v+i)(ak-v+k+i)}
> \!\int_{1}^{a+1} \hspace{-1em} g(t)\,dt + \frac{g(1){-}g(a+1)}{2},
\end{align*}
\stopcompact{small}
which implies that
\startcompact{small}
\begin{align*}
& \Delta(a,v,k)
> \ln\left( \frac{(ak-v+a+1)(ak-v+k+1)}{(ak-v+k+a+1)(ak-v+1)} \right)
\\*
&\hspace{5.3em} + \frac{k}{2(ak-v+1)(ak-v+k+1)}
\\*
&\hspace{5.3em} - \frac{k}{2(ak-v+a+1)(ak-v+k+a+1)}
\\*
&\hspace{5.3em} - \frac{1}{ak-v+k+a+1} \quad
\triangleq\quad h(a,v,k).
\end{align*}
\stopcompact{small}
Now, it suffices to show that $h(a,v,k)\geq 0$ for any \mbox{$a,v,k\in\ZZ^+$}, \mbox{$k\geq\frac{v}{a}$}.
This is indeed the case since
\[
\lim_{k\rightarrow\infty} h(a,v,k) = 0,
\]
and the partial derivative $\frac{\partial}{\partial k} h(a,v,k)$, which is given by
\startcompact{small}
\begin{align*}
& \frac{a}{2} \left( \frac{2(ak-v+a+1)+1}{(ak-v+a+1)^{2}} - \frac{2(ak-v+1)+1}{(ak-v+1)^{2}} \right) \quad +
\\*
&\; \frac{a+1}{2} \left( \frac{2(ak-v+k+1)+1}{(ak-v+k+1)^{2}} - \frac{2(ak-v+k+a+1)-1}{(ak-v+k+a+1)^{2}} \right) \!,
\end{align*}
\stopcompact{small}
can be shown to be negative.
\hfill~
\end{proof}

It follows from Lemma~\ref{thm:lemma:ExpectedDelaySymmetricFixedV} that for each \mbox{$v\in\ZZ^+$}, the expected recovery delay $\EDm{\floor{kT}}$ decreases as $k$ takes larger values in the interval \mbox{$\big((v-1)\ell,v\ell\big]$}, that is,
\begin{align*}
&\quad\;\; \EDm{\floor{\big(\floor{(v-1)\ell}+1\big)T}}
\\
&\; > \EDm{\floor{\big(\floor{(v-1)\ell}+2\big)T}}
\\
&\; > \cdots
\\
&\; > \EDm{\floor{\floor{v\ell}T}}.
\end{align*}
We will proceed to show that
\[
\EDm{\floor{\floor{v\ell}T}}
\geq \EDm{\floor{\floor{\ell}T}}
\]
for all \mbox{$v\in\ZZ^+$}.
This is equivalent to showing that
\[
\sum_{i=1}^{\floor{v\ell}} \frac{1}{a\floor{v\ell}-v+i}
\geq \sum_{i=1}^{\floor{\ell}} \frac{1}{a\floor{\ell}-1+i}
\]
for any \mbox{$\ell\geq 1$}, \mbox{$a,v\in\ZZ^+$}.
According to Lemma~\ref{thm:lemma:ExpectedDelaySymmetricFixedV}, we have
\[
\sum_{i=1}^{\floor{v\ell}} \frac{1}{a\floor{v\ell}-v+i}
\geq \sum_{i=1}^{v\floor{\ell}+v-1} \frac{1}{a\left(v\floor{\ell}+v-1\right)-v+i},
\]
since we can substitute $\ell$ with \mbox{$\floor{\ell}+\tau$}, where \mbox{$\tau\in[0,1)$}, which yields
\[
\floor{v\ell} = \floor{v\floor{\ell} + v\tau} = v\floor{\ell} + \floor{v\tau}
\leq v\floor{\ell}+v-1.
\]
Defining the function
\begin{align*}
f(a,\ell,v)
&\triangleq \sum_{i=1}^{v\ell+v-1} \frac{1}{a\left(v\ell+v-1\right)-v+i}
\\
&= H_{\big((a+1)(\ell+1)-1\big)v-(a+1)} - H_{\big(a(\ell+1)-1\big)v-a},
\end{align*}
it therefore suffices to show that
\begin{align}
f(a,\ell,v) \geq f(a,\ell,v{=}1) \label{eq:ExpectedDelaySufficientCondition}
\end{align}
for any \mbox{$a,\ell,v\in\ZZ^+$}.

To obtain lower and upper bounds for $f(a,\ell,v)$, we apply the following bounds for the harmonic number $H_n$, \mbox{$n\geq 1$} \cite{dl:havil03exploring}:
\startcompact{small}
\begin{align*}
\underbrace{\ln\left(\!n+\frac{1}{2}\right)\!+\gamma+\frac{1}{24(n+1)^{2}}}_{\triangleq H_{\text{LB}}(n)}
< H_n <
\underbrace{\ln\left(\!n+\frac{1}{2}\right)\!+\gamma+\frac{1}{24n^{2}}}_{\triangleq H_{\text{UB}}(n)},
\end{align*}
\stopcompact{small}
where $\gamma$ is the Euler-Mascheroni constant.
This produces the lower bound
\begin{align*}
f_{\text{LB}}(a,\ell,v)
&\triangleq H_{\text{LB}}\left(\big((a+1)(\ell+1)-1\big)v-(a+1)\right)
\\*
&\quad - H_{\text{UB}}\left(\big(a(\ell+1)-1\big)v-a\right),
\end{align*}
and the upper bound
\begin{align*}
f_{\text{UB}}(a,\ell,v)
&\triangleq H_{\text{UB}}\left(\big((a+1)(\ell+1)-1\big)v-(a+1)\right)
\\*
&\quad - H_{\text{LB}}\left(\big(a(\ell+1)-1\big)v-a\right),
\end{align*}
for $\big(a(\ell+1)-1\big)v-a\geq 1$.
The lower bound $f_{\text{LB}}(a,\ell,v)$ is an increasing function of $v$ for any \mbox{$a\geq 1$}, \mbox{$\ell\geq 1$}, \mbox{$v\geq 2$}, since the partial derivative $\frac{\partial}{\partial v} f_{\text{LB}}(a,\ell,v)$,
which is given by
\startcompact{footnotesize}
\begin{align*}
& \frac{2(\ell-1)}{\left(2\big((a+1)(\ell+1)-1\big)v-2(a+1)+1\right) \left(2\big(a(\ell+1)-1\big)v-2a+1\right)}
\\*
&\qquad + \frac{a(\ell+1)-1}{12\left(\big(a(\ell+1)-1\big)v-a\right)^{3}}
- \frac{(a+1)(\ell+1)-1}{12\left(\big((a+1)(\ell+1)-1\big)v-a\right)^{3}},
\end{align*}
\stopcompact{footnotesize}
can be shown to be positive.
We therefore have
\[
f(a,\ell,v)
\geq f_{\text{LB}}(a,\ell,v)
\geq f_{\text{LB}}(a,\ell,v{=}2)
\]
for any \mbox{$v\geq 2$}, \mbox{$a,\ell,v\in\ZZ^+$}.
We now proceed to demonstrate that \mbox{$f_{\text{LB}}(a,\ell,v{=}2) \geq f(a,\ell,v{=}1)$}.

For the case $\ell=1$, consider the function
\begin{align*}
g(a)
&\triangleq f_{\text{LB}}(a,\ell{=}1,v{=}2) - f(a,\ell{=}1,v{=}1)
\\
&= \ln\left(\frac{2a+1}{2a-1}\right) - \frac{81a^4-71a^2+16}{a(9a^2-4)^{2}}.
\end{align*}
It suffices to show that \mbox{$g(a)\geq 0$} for any {$a\geq 1$}, which is indeed the case since
\[
\lim_{a\rightarrow\infty} g(a) = 0,
\]
and the derivative
\[
g'(a) = -\frac{621a^6-961a^4+436a^2-64}{a^2(4a^2-1)(9a^2-4)^{3}}
\]
is negative.

For the case $\ell\geq 2$, we consider the function
\[
h(a,\ell)
\triangleq f_{\text{LB}}(a,\ell,v{=}2) - f_{\text{UB}}(a,\ell,v{=}1),
\]
which can be shown to be nonnegative for any $a\geq 1$, $\ell\geq 2$.
It follows that
\[
f_{\text{LB}}(a,\ell,v{=}2)
\geq f_{\text{UB}}(a,\ell,v{=}1)
\geq f(a,\ell,v{=}1)
\]
for any \mbox{$\ell\geq 2$}, \mbox{$a,\ell\in\ZZ^+$}.

Combining these results, we obtain
\[
f(a,\ell,v)
\geq f_{\text{LB}}(a,\ell,v)
\geq f_{\text{LB}}(a,\ell,v{=}2)
\geq f(a,\ell,v{=}1)
\]
for any \mbox{$v\geq 2$}, \mbox{$a,\ell,v\in\ZZ^+$}, which gives us inequality~\eqref{eq:ExpectedDelaySufficientCondition} as required.
Consequently, we have
\[
\EDm{\floor{kT}}
\geq \EDm{\floor{\floor{\ell} T}}
\]
for any $k\in\ZZ^+$.
Since
\startcompact{small}
\begin{align*}
\EDm{n}
\begin{cases}
= \EDm{\floor{\floor{\frac{n}{T}}T}} & \text{if } \frac{n}{T}\in\ZZ^+,
\\
\geq \EDm{\floor{\left(\floor{\frac{n}{T}}+1\right)T}} & \text{otherwise},
\end{cases}
\end{align*}
\stopcompact{small}
we also have
\[
\EDm{n}
\geq \EDm{\floor{\floor{\ell}T}}.
\]
Therefore, if $\floor{\ell}\leq\floor{\frac{n}{T}}$, then
\xxsm{\floor{\floor{\ell}T}} is an optimal symmetric allocation.
On the other hand, if $\floor{\ell}>\floor{\frac{n}{T}}$, then we can eliminate all but the two largest candidate values for $m^*$ in \eqref{eq:CandidateMs}, since
\begin{align*}
& \EDm{\floor{T}}
> \EDm{\floor{2T}}
> \cdots
\\*
&\hspace{13em} > \EDm{\floor{\floor{\frac{n}{T}}T}}
\end{align*}
by Lemma~\ref{thm:lemma:ExpectedDelaySymmetricFixedV}.
\hfill~
\end{proof}

\StopLongVersion

\bibliographystyle{IEEEtran}
\bibliography{IEEEabrv,dleong}

\end{document}